\documentstyle[aps,prd,epsf]{revtex}
\begin{document}
\hspace{-10mm}
\leftline{\epsfbox{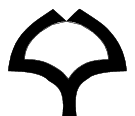}}
\vspace{-10.0mm} % for revtex
\thispagestyle{empty}
{\baselineskip-4pt
\font\yitp=cmmib10 scaled\magstep2
\font\elevenmib=cmmib10 scaled\magstep1  \skewchar\elevenmib='177
\leftline{\baselineskip20pt
%\hspace{8mm} % for revtex
\hspace{10mm} % for revtex
\vbox to0pt
   { {\yitp\hbox{Osaka \hspace{1.5mm} University} }
     {\large\sl\hbox{{
Theoretical Astrophysics}} }\vss}}

\rightline{\large\baselineskip20pt\rm\vbox to20pt{
\baselineskip14pt
\hbox{OU-TAP-180}
%\hbox{gr-qc/xxx}
\vspace{1mm}
%\hbox{October 2000}\vss}}%
\hbox{\today}\vss}}%
}
\vskip8mm
\begin{center}{\large \bf
Casimir energy for de Sitter branes in bulk AdS$_5$}
\end{center}
%\vspace*{2mm}
%\centerline{\large \bf
%}
\vspace*{4mm}
\centerline{\large
Wade Naylor,\footnote{E-mail:wnaylor@vega.ess.sci.osaka-u.ac.jp} 
Misao Sasaki\footnote{E-mail:misao@vega.ess.sci.osaka-u.ac.jp}}
\vspace*{4mm}
\centerline{\em Department of Earth and Space Science, Graduate
School of Science,}
\centerline{\em Osaka University, Toyonaka 560-0043, Japan}
%%%%%%%%%%%%%%%%%%%%%%%%%%%%%%%%%%%%%%%%%%%%%%
\vspace*{4mm}

\begin{abstract}
The vacuum energy for a massless conformally coupled scalar field
in a brane world set up,
corresponding to de Sitter branes in a bulk anti-de Sitter spacetime, 
is calculated. We use the
Euclidean version of the metric which can be conformally
related to a metric similar in form to the
Einstein universe ($S^4\times R$). Employing $\zeta$-function
regularisation we evaluate the
one-loop effective potential and show that the vacuum energy is
zero for the one brane and
non-zero for the two brane configuration. We comment on the
back-reaction of this Casimir energy and on the inclusion of a mass term 
or non-conformal coupling.
\vspace*{4mm}

PACS number(s): 04.50.+h, 11.25.Mj, 98.80.Cq
\end{abstract}
\pacs{PACS number(s): 04.50.+h, 11.25.Mj, 98.80.Cq}

%\vspace*{3mm}
Many ideas and proposals within the circumference of
what we call brane world cosmology
(BWC) relate to the work of Randall and Sundrum (RS) \cite{RS1}
 who initially implemented their
two brane model to try and solve the hierarchy problem.
 In a variant of this model (a single
brane with positive tension \cite{RS2}) they also showed
 that the extra dimension could be
of infinite extent. Generally speaking, the second model
 (and its extensions to include curved
branes) has been of more interest to the cosmology
 community because gravity behaves
essentially like 4-dimensional gravity localised
 near the brane\cite{RS2,SMS,GT}.

An interesting BWC scenario has been developed in \cite{HIME},
 where it is possible to obtain
inflation on a single positive tension brane solely due
 to the effect of a  bulk gravitational scalar field.
In this regard, the vacuum energy of the bulk scalar field
could have some affect on the cosmological evolution of the
brane, (depending on the size and sign of this quantity) and
is an interesting subject in its own
right. The importance of vacuum energy in brane world
 scenarios (BWS) has also been
highlighted in \cite{MUKO} where it has been suggested that
it may be possible to solve the
hierarchy problem with two positive tension branes, if
the back-reaction of the Casimir effect is
included (also see \cite{HKP}).

In this letter we evaluate the one-loop effective potential
 (leading to the Casimir energy) for one and
two de Sitter brane configurations in a bulk 5-dimensional
anti-de Sitter spacetime. A massless
conformally coupled scalar field is investigated as
this is the simplest case technically.
In \cite{NOZ} a similar calculation has been considered
for spherical and hyperbolic one brane
configurations. However, we consider a spherical
two brane configuration and for the one brane
case the results agree. Quite extensive work has been done 
for flat branes and we refer the reader to the literature, see 
\cite{TOMS,GPT,GR,FT,FMT,FMT2} and the references therein.

Calculations of the Casimir energy are non-intuitive in how the 
overall sign depends on the topology and/or dimension of the 
spacetime \cite{MILT}.
One has to perform the calculation. To obtain the result, 
we employ the
$\zeta$-function regularisation method \cite{TEN}.
The beauty of $\zeta$-function regularisation
(or some would say its weakness) is that it automatically
removes the divergences of the theory
leaving a finite result. An interesting discussion about
different analytic continuations in
$\zeta$-function regularisation can be found in \cite{SVAI}.

To perform our calculation we use the Euclideanised form of
the metric suitable for de Sitter branes \cite{GS}.
 It can be written in conformal form as
\begin{equation}\label{metric}
ds^2 =a^2(z)(dz^2+d\chi^2 +\sin^2\chi d\Omega_{(3)}^2)\,;
\quad a(z)=\frac{\ell}{\sinh(z_0+|z|)}\,,
\end{equation}
where $\ell=(-6/\Lambda_5)^{1/2}$ is the anti-de Sitter radius,
$d\Omega_{(3)}^2$ is the metric
on the 3-sphere and the positive tension brane is placed at $z=0$.
Here, we place the other negative tension brane at $|z|=L$
 (the one brane configuration is given by $L\rightarrow\infty$).
 The brane tensions in this case are \cite{GS,GEN}
\begin{equation}
\sigma_+=\frac{3}{4\pi G_5 \ell}\cosh z_0,
\qquad\sigma_-=-\frac{3}{4\pi G_5 \ell}\cosh(z_0+L),
\end{equation}
and the non-dimensional length $L$ is given in terms of the 
physical length $r$ as
\begin{eqnarray}
L=\int_{r_-}^{r_+}\frac{dr}{a}=\int_{r_-}^{r_+}\frac{dr}
{\ell\sinh(r/\ell)}\,.
\end{eqnarray}
The metric is conformal to the Euclidean version of an
Einstein universe $(S^4\times R)$ with
unit radius. The topology of this metric corresponds to two concentric
hyperspheres of equal radius on a cylinder of length $L$
 (the conformal scaling has shrunk
 the radii of the two hyperspheres to equal radius).

Because we are dealing with a conformally coupled scalar field we must

find the eigenvalues of the Klein-Gordon equation given by
(e.g., see \cite{BD})
\begin{eqnarray}
\left (-\partial_ z{}^2-\Delta^{(4)}+\xi_5R^{(4)}\right)\phi&=&\lambda^2\phi,
\nonumber\\
\label{eigen}
\left [\left(\frac{\pi n}{L}\right)^2-\Delta^{(4)}+\frac{9}{4}\right]\phi
&=&\lambda^2\phi,
\end{eqnarray}
where $\xi_5=3/16$, $R^{(4)}=12$ and imposing $Z_2$ symmetry on the branes
implies Neumann boundary conditions giving the
discrete sum between the brane boundaries,
for a conformally coupled scalar field \cite{GPT}.
In general, as pointed out in \cite{NOZ},
the eigenfunctions will satisfy mixed boundary conditions
and the spectrum of $-\partial_z{}^2$ will not be easily obtained.
For the operator $-\Delta^{(4)}+9/4$,
the eigenvalues are (for Gegenbauer polynomials see \cite{BATE}.)
\begin{equation}
\lambda_m^2=(m+3/2)^2,\qquad\qquad d(m)=
\frac{1}{3}(m+2)(m+3/2)(m+1),
%\frac{1}{3}\left((m+3/2)^3-\frac{1}{4}(m+3/2)\right),
\label{deg}
\end{equation}
where $d(m)$ is the degeneracy.
Thus the eigenvalues for the Klein-Gordon operator are
\begin{eqnarray}
\lambda_{n,m}^2=\left(\frac{\pi n}{L}\right)^2+(m+3/2)^2\,.
\end{eqnarray}
%rearranged for later convenience.
Employing the Euclidean effective action, we define the function
\begin{equation}
\label{gfunc}
g(s)=\sum_{m,n=0}^\infty d(m) \lambda_{m,n}^{-2s}.\
\end{equation}
and then the one-loop effective action, $W_1$, is related to $g(s)$ by
\begin{equation}
\label{effact}
W_1=-\frac{1}{2}\,g'(0).
\end{equation}
Note, there is no renormalisation scale, $\mu$,
 in the function $g(s)$ because the eigenvalues are
dimensionless (due to the conformal rescaling).

We first begin with the simpler case of the vacuum energy for the
Einstein universe $S^4\times R$.
This is equivalent to finding the zero point energy of the equation
\begin{equation}
\partial_t^2\psi-\Delta^{(4)}\psi+\frac{3}{16}R^{(4)}\psi=0.
\end{equation}
The eigenvalues and degeneracy are still given by Eq. (\ref{deg}).
 It is then straightforward to show that
\begin{equation}
E_0=\frac{1}{2}\,g(-1/2)=
\frac{1}{6}\left(\zeta(2s-3,3/2)
-\frac{1}{4}\zeta(2s-1,3/2)\right)_{s\rightarrow -1/2}=0,
\label{ein}
\end{equation}
where $\zeta(a,b)$ is the generalised (or Hurwitz)
 $\zeta$-function and is related to
the Bernoulli polynomial $B_n(x)$ by \cite{BATE}
\begin{equation}
\zeta(-n,a)=-\frac{B_{n+1}(a)}{n+1}\;.
\label{bern}
\end{equation}
It is easy to verify that $E_0$ is zero, using Eq. (\ref{bern}),
as should be expected
because in an odd number of dimensions there is no conformal
anomaly \cite{BD}. This calculation
is similar to the one by Ford \cite{FORD} for $S^3\times R$,
where an exponential cut off was
used in the mode sum rather than $\zeta$-function
regularisation (for $S^2\times R$ see \cite{BO}).

We now evaluate the effective potential for a one brane configuration,
where $L\rightarrow\infty$ in
Eq. (\ref{eigen}) and the discrete sum over $n$ is now an integral.
The function $g(s)$ becomes
\begin{equation}
g_1(s)=\frac{2L}{\pi}\int_0^\infty dk \sum_{m=0}^\infty d(m)
\left(k^2+(m+3/2)^2\right)^{-s},
\end{equation}
where the factor of $2$ is because there are two copies of the bulk space on
either side of the brane. For large $s$ we can interchange
the order of the sum and the integral
and perform the $k$ integration, leading to
\begin{eqnarray}
g_1(s)&=&\frac{2L}{\pi}\,
\frac{\sqrt{\pi}}{2}\sum_{m=0}^\infty
\frac{\Gamma(s-1/2)}{\Gamma(s)}\;d(m)(m+3/2)^{1-2s},
\nonumber\\
&=&\frac{L}{\pi}\frac{\sqrt{\pi}}{3}\frac{\Gamma(s-1/2)}{\Gamma(s)}
\left(\zeta(2s-4,3/2)-\frac{1}{4}\zeta(2s-2,3/2)\right),
\end{eqnarray}
where in the second step we have used simple algebra to
rewrite the equation in terms of
generalised $\zeta$-functions. For $s=0$ it is clear
that $g_1(0)=0$ because
\begin{equation}
\label{gam}
\frac{1}{\Gamma(s)}=s+\gamma s^2+O(s^3),
\end{equation}
where $\gamma$ is Euler's constant. Thus,
\begin{equation}
\label{1bravac}
g_1'(0)=\frac{L}{\pi}\frac{\sqrt{\pi}}{3}\Gamma(-1/2)
\left(\zeta(-4,3/2)-\frac{1}{4}\zeta(-2,3/2)\right)=0,
\end{equation}
which is zero because the combination of $\zeta$-functions
in Eq. (\ref{1bravac}) cancel, as can be
verified by employing Eq. (\ref{bern}).
Therefore, the one-loop effective potential
(and hence the vacuum energy) for a conformally
coupled scalar field is zero for the one brane case and
does not affect BWC models, such as in
\cite{HIME}. This is consistent with the result found in \cite{NOZ}.
%This may not be the case for other fields, as pointed out in \cite{NOZ}.
Although, one might naively expect imposing a boundary on the brane
similar to that for a half Einstein universe
will give a non-zero contribution.

Now we come to the main task at hand. The two brane configuration
has two infinite summations, which can be expressed in
terms of generalised $\zeta$-functions. In fact,
there are many subtle issues regarding the correct analytic
continuation of such functions
upon interchange of the order of the summations
(see \cite{TEN} for a detailed discussion).
The function $g(s)$ becomes
\begin{eqnarray}
g_2(s)&=& \sum_{n=-\infty}^\infty\,
\sum_{m=0}^\infty d(m)\left(c^2 n^2+(m+3/2)^2\right)^{-s},
\nonumber\\
\label{2bravac}
&=&\frac{1}{\Gamma(s)} \sum_{n=-\infty}^\infty\,
\sum_{m=0}^\infty d(m)\int_0^\infty
dt\;t^{s-1} \exp\{-t[\;c^2n^2+(m+3/2)^2]\},
\end{eqnarray}
where $c=\pi/L$ and in the second step we have made use
of the Mellin transform.
The sum over $n$ is from $-\infty$ to $\infty$ because
imposing orbifold boundary conditions
essentially doubles the modes.
This formula is almost identical in form to the
2-dimensional Epstein-Hurwitz $\zeta$-function
studied by Elizalde in \cite{ELI2} (see also \cite{TEN}),
apart from the degeneracy factor $d(m)$.
This work has many interesting results and the one we employ is
\begin{equation}
\sum_{n=-\infty}^\infty\;e^{-c^2n^2}
=\sqrt{\frac{\pi}{c^2}} +2\sqrt{\frac{\pi}{c^2}}
\sum_{n=1}^\infty\,\exp[-\pi^2n^2/c^2].
\label{sum}
\end{equation}
This is non other than the unimodular transformation formula of 
Jacobi's $\theta_3$ function \cite{BATE} (the Poisson resummation formula). 
Substitution of the above equation into (\ref{2bravac})
allows us to interchange the order of the summations.
After integrating with respect to $t$ and
performing some simple algebra we obtain
\begin{eqnarray}
g_2(s)&=&
+\frac{1}{3}\sqrt{\frac{\pi}{c^2}}\frac{\;\Gamma(s-1/2)}{\Gamma(s)}
\left(\zeta(2s-4,3/2)-\frac{1}{4}\zeta(2s-2,3/2)\right)
\nonumber\\
&&+\frac{4\pi^s}{\Gamma(s)}\,c^{-s-1/2}\sum_{n=1,\;m=0}^\infty
\;d(m)\;n^{s-1/2}(m+3/2)^{-s+1/2}\,
K_{s-1/2}[2\pi c^{-1}n(m+3/2)].
\end{eqnarray}
It is easy to verify that $g_2(0)=0$, using Eq. (\ref{gam}).
Furthermore, taking the derivative of $g_2(s)$ with respect to $s$
(and leaving only terms that
remain independent of $\Gamma(s)$ near $s=0$, see Eq. (\ref{gam})) gives
\begin{eqnarray}
g_2'(0)&=&
+\frac{1}{3}\sqrt{\frac{\pi}{c^2}}
\Gamma(-1/2)\left(\zeta(-4,3/2)-\frac{1}{4}\zeta(-2,3/2)\right)
\nonumber\\
&&+4c^{-1/2}\sum_{n=1,\;m=0}^\infty
\;d(m)\;n^{-1/2}(m+3/2)^{1/2}\;K_{-1/2}[2\pi c^{-1}n(m+3/2)],
\end{eqnarray}
where a prime denotes differentiation with respect to $s$.
Interestingly, the first term is exactly the contribution from the
one brane configuration
Eq. (\ref{1bravac}), which is zero. This is similar to finite temperature
field theory
where the $n=0$ mode gives the zero temperature contribution
(this can be clearly
seen by using heat kernel methods \cite{WN}, see also \cite{NOZ}).
Of course, this kind of
behaviour was expected due to the topology of the conformally
re-scaled metric, Eq. (\ref{metric}).
%The above terms can be related to the heat kernel of the operator!!!
The second term is non-zero and depends on the value of $c=\pi/L$.
Note that $K_{-1/2}(z)=\sqrt{\pi/(2z)}\,e^{-z}$.
Therefore, the one-loop effective potential per unit volume is given by,
after some simplification,
\begin{equation}
V_1=\frac{W_1}{2L{\rm Vol(S^4)}}=-\frac{A(L)}{8\pi^2L}\,,
\end{equation}
where $W_1$ is the effective action for the conformally re-scaled field and
\begin{eqnarray}
A(L)&=&
\sum_{n=1,\;m=0}^\infty\;\frac{(m+2)(m+3/2)(m+1)}{n}\,\exp[-2Ln(m+3/2)]
\nonumber\\
&=&-\sum_{m=0}^\infty\;(m+2)(m+3/2)(m+1)\,\ln(1-\exp[-L(2m+3)])\,.
\end{eqnarray}

The function $A(L)$ is generally a complicated function of $L$, but
its qualitative behaviour can be deduced analytically. First, it
is easy to see that $A(L)>0$ and $dA(L)/dL<0$ for $L>0$.
Note, that we have normalized the radius of S$^4$ to unity. Therefore,
the limit $L\to 0$ corresponds to infinitely large 4-spheres on the
boundaries, i.e., flat space.
In the limit $L\ll1$, we may approximate $A(L)$ as
\begin{eqnarray}
A(L)
&\approx& -\int_0^\infty dm\, m^3\ln(1-\exp[-2mL])
\nonumber\\
&=&\frac{1}{4(2L)^4}\int_0^\infty dx\frac{x^4}{e^x-1}
=\frac{\Gamma(5)\zeta(5)}{4(2L)^4}\,,
\end{eqnarray}
which gives
\begin{equation}
V_1\approx-\frac{\Gamma(5)\zeta(5)}{2^{10}\pi^2L^5}
\quad\mbox{for}\quad L\ll1\,.
\label{smallL}
\end{equation}
Using the relation,
\begin{eqnarray}
\zeta'(-2n)=(-1)^n\frac{\Gamma(2n+1)\zeta(2n+1)}{2(2\pi)^{2n}}
\quad(n=1,2,3,\cdots)\,,
\end{eqnarray}
we find Eq.~(\ref{smallL}) exactly agrees with the flat space Casimir
energy for a conformal scalar \cite{GPT}.
It may be worth mentioning that $V_1$ given by Eq.~(\ref{smallL})
is also equal to the free energy of a massless boson
at temperature $T=1/(2L)$, except for the factor of 2 that comes
from the mode doubling.
In the limit $L\gg1$, $A(L)$ behaves as
\begin{eqnarray}
A(L)\approx \sum_{m=0}^\infty\;(m+2)(m+3/2)(m+1)\,\exp[-L(2m+3)]
\sim 3\exp[-3L]\,.
\label{largeL}
\end{eqnarray}
Thus, the Casimir energy for de Sitter boundaries decreases much more
rapidly than that for flat space boundaries as the non-dimensional
length $L$ increases.
We have checked our results for the sum $A(L)$
on different numerical packages and find that, for example,
when $L=0.5$, $1$, and $2$,
$A(L)=5.90$, $0.31$ and $0.008$, respectively.

From the symmetries of the metric, Eq.~(\ref{metric}),
and because the stress energy tensor for a
conformally coupled scalar field is traceless in odd dimensions
(no conformal anomaly, see \cite{BD})
the Casimir energy density, $\rho$, is equal to the effective potential, $V_1$.
Also, using the fact that the conformally re-scaled stress energy
tensor is given in terms of the
original one by $\left<T^\mu{}_\nu\right>_g=a^{-D}\left<T^\mu{}_\nu\right>$,
see \cite{BD,GPT},
we have
\begin{equation}
\left<T^\mu{}_\nu\right>_g
=\ell^{-5}\sinh^5(z_0+|z|)\left<T^\mu{}_\nu\right>.
\end{equation}
This then implies that the physical
energy density $\rho_g$ is given by
\begin{equation}
\label{enden}
\rho_g=-\frac{\sinh^5(z_0+|z|)}{8\pi^2 \ell^5 L}\,A(L)\,.
\end{equation}
We now discuss the implications of these analyses on the issues raised 
in the introduction.

The back-reaction of the Casimir energy plays an important role
in BWS \cite{MUKO,HKP}, including
inflating branes. However, in \cite{MUKO} the back-reaction for
the de Sitter brane model
only considered the vacuum energy from Minkowski branes.
 For flat branes this only depends on
the difference between the number of bosons ($N_b$) and fermions ($N_f$),
where $N_b=N_f$
gives the usual RS model and $N_b>N_f$ gives two positive tension branes,
 see \cite{MUKO}.
Thus, we should also consider other fields such as the fermion.
However, in contrast to Minkowski
branes, the eigenvalues in the mode sum
for two de Sitter branes are different for bosons and fermions,
 see \cite{FORD} for $S^3\times R$,
implying the vacuum energy will not be of equal magnitude,
 but opposite in sign. Therefore, we can
infer, from the previous sentences, that the bulk will not remain
pure anti-de Sitter under the back-reaction
 (for de Sitter branes) even for $N_b=N_f$.

In the context of BWC models, such as in \cite{HIME},
we must consider the more general
case of a massive or non-conformal scalar field.
Adding a mass term will break conformal invariance.
Altering the conformal coupling constant
$\xi_5$ not only breaks conformal
invariance, but also modifies the boundary conditions.
As we mentioned earlier (also see \cite{NOZ}), the
boundary conditions will become mixed.
One can of course consider a small perturbation
of $\xi_5$ or a small mass term. Such cases may be
treated analytically.
Otherwise, the eigenspectrum of $-\partial_z^2$
needs to be solved numerically.
Nevertheless, for large $n$ the spectrum
will return to $\pi n/L$ and we expect to need only evaluate
this correction to the Casimir energy up
to, say, $n\approx 10$.

In forthcoming work \cite{WN} we will present a more detailed
discussion of the $\zeta$-function regularisation procedure.
There, its relation to the heat kernel will be made explicit.
In connection with the generalisation to non-conformal cases,
we mention the work done by Dowker and Apps \cite{DA} for 
generalised cylinders (up to 4-dimensions), and the work by
Garriga, Pujol\`{a}s and Tanaka\cite{GPT,GPT2} for 
a 5-dimensional spacetime bounded by flat branes.
They stress the importance of the cocycle function.
Also, we will present the results for a spinor field 
and an analysis of the introduction of
a small mass term (or non-conformal coupling) into the wave equation.
Other avenues of research for de Sitter branes include the
general non-conformal case, finite temperature effects
(see \cite{BREV} for the RS model and also the comments made in \cite{ROTH}), 
boundaries on the brane, and the possibility of stabilising the radion field.
%Other methods of comparison should also be used to check our results, 
%such as the Green's function method.

%%%%%%%%%%%%%%%%%%%%%%%%%%%%%%%%%%%%%%%%%%%%%
\acknowledgements
W.N. acknowledges support from JSPS for Postdoctoral Fellowship for Foreign Researchers 
No. P01773.
The work of M.S. is supported by Monbukagaku-sho Grant-in-Aid
for Scientific Research (S) No. 14102004.
We are grateful to Ian Moss for useful discussions 
and drawing our attention to 
reference \cite{DA}. 
We would also like to thank Yoshiaki Himemoto, Alan Knapman and 
Takahiro Tanaka for interesting conversations.

%%%%%%%%%%%%%%%%%%%%%%%%%%%%%%%%%%%%%%%%%%%%%

%%%%%%%%%%%%%%%%%%%%%%%%%%%%%%%%%%%%%%%%%%%%%%

\end{document}